\documentclass[11pt]{article}

\usepackage[a4paper, margin=2.5cm]{geometry}
\usepackage{setspace}
\usepackage{amsmath,amssymb,mathtools}
\usepackage{bm}
\usepackage{nicefrac}

\usepackage{graphicx}
\usepackage[caption=false]{subfig}
\usepackage{booktabs}
\usepackage{array}
\usepackage{multirow}
\usepackage{adjustbox}
\usepackage{makecell}

\usepackage{enumerate}
\usepackage{color,xcolor}
\usepackage{newtxtext,newtxmath}
\usepackage[affil-it]{authblk}

\usepackage[numbers,sort&compress]{natbib}
\usepackage[colorlinks=true,linkcolor=blue,citecolor=blue,urlcolor=blue]{hyperref}
\usepackage{xurl}
\usepackage{cleveref}

\begin{document}
% \linenumbers

\title{Multi-Mode Quantum Annealing for Generative Representation Learning with Boltzmann Priors}

\author[1]{Gilhan Kim}
\author[1,2,3]{Daniel K. Park\thanks{Corresponding author: dkd.park@yonsei.ac.kr}}

\affil[1]{Department of Statistics and Data Science, Yonsei University, Seoul 03722, Republic of Korea}
\affil[2]{Department of Applied Statistics, Yonsei University, Seoul 03722, Republic of Korea}
\affil[3]{Department of Quantum Information, Yonsei University, Seoul 03722, Republic of Korea}

\date{}
\maketitle
\vspace{-1.2cm}
%%%%%%%%%%%%%%%%%%%%%%%%%%%%%%%%%%%%%%%%%%%%%%%%%%%%%%%%%%%
\begin{abstract}
Energy-based models provide a natural bridge between statistical physics and machine learning by representing data through structured energy landscapes. Boltzmann machines are a particularly compelling class of such models for capturing complex interactions among latent variables, but their use in modern generative learning has been limited by the classical intractability of sampling from general (non-restricted) Boltzmann distributions. Here we develop a quantum-annealing-based framework that enables variational autoencoders with general Boltzmann priors. The framework employs three complementary annealing modes tailored to different stages of learning and deployment: diabatic quantum annealing provides unbiased Boltzmann samples for efficient training, slower annealing concentrates samples near low-energy configurations of the learned prior for unconditional generation, and conditional annealing with external fields steers the learned energy landscape toward attribute-specific regions for conditional generation and semantic editing. Using up to 2000 qubits on a D-Wave Advantage2 processor, we demonstrate stable training and high-quality generation on MNIST, Fashion-MNIST, and CelebA, achieving faster convergence and lower reconstruction loss than a Gaussian-prior VAE with the same encoder–decoder architecture. Beyond generation, the learned energy function provides out-of-distribution detection signals that add discriminative power beyond reconstruction loss. We demonstrate that these scores separate in-distribution samples from held-out digit classes in one-class MNIST experiments and improve the detection of market regime shifts in financial data. These results establish quantum annealing as a practical and controllable physical mechanism for energy-based representation learning and generative modeling beyond the reach of tractable classical approaches.
\end{abstract}
%%%%%%%%%%%%%%%%%%%%%%%%%%%%%%%%%%%%%%%%%%%%%%%%%%%%%%%%%%%

%%%%%%%%%%%%%%%%%%%%%%%%%%%%%%%%%%%%%%%%%%%%%%%%%%%%%%%%%%%
\section*{Introduction}
%%%%%%%%%%%%%%%%%%%%%%%%%%%%%%%%%%%%%%%%%%%%%%%%%%%%%%%%%%%

Learning structured, low-dimensional representations from high-dimensional data is a central problem in machine learning, with applications spanning generative modeling, scientific discovery, and data-driven decision-making.
Variational autoencoders (VAEs) have become a standard framework for addressing this task by jointly learning an encoder that maps high-dimensional observations to a compact latent space and a decoder that reconstructs or generates data from latent variables \cite{kingma2013auto,rezende2014stochastic}.
A central design choice in VAEs is the latent prior, which shapes the structure of the learned representation and strongly influences both learning dynamics and downstream performance.
In most practical settings, the prior is chosen to be factorized, typically an isotropic Gaussian distribution, due to its analytical convenience and stable optimization. However, this simplicity comes at a cost: factorized priors impose independence among latent variables and therefore limit the ability of the latent space to represent structured interactions, correlations, and collective modes of variation that may be important for downstream generation and inference.

A natural way to move beyond this limitation is to replace the factorized prior with an energy-based model \cite{lecun2006tutorial}, which bridges statistical physics and machine learning by representing latent structure through an energy function. Among such models, Boltzmann machines (BMs) are particularly attractive because they define probability distributions over binary latent variables through learned pairwise interactions, enabling highly structured dependencies beyond factorized priors \cite{ackley1985learning,sussmann1988learning,younes1996synchronous}.

When used as a prior in a VAE, a BM allows the latent space to be shaped by learned interactions rather than by a fixed parametric assumption. This is particularly important for generation: whereas a factorized prior samples latent variables independently, a Boltzmann prior couples them so that sampling naturally favors coherent latent configurations. Moreover, the explicit energy function endows the latent space with an energy-landscape structure, providing a physically interpretable perspective on learning and enabling control mechanisms tailored to different learning and generation tasks (Fig.~\ref{fig:sampling}). In principle, this can substantially enrich the representational and generative capacity of the model.

However, this greater expressivity comes at a fundamental computational cost: the normalization constant is generally intractable, preventing exact evaluation of expectations under the prior and requiring sampling-based methods during training and generation.
In classical settings, feasible sampling typically requires either restricting the model structure or changing the latent-space formulation. Discrete VAEs, for example, have employed restricted Boltzmann machine priors, whose bipartite structure permits efficient block Gibbs sampling~\cite{rolfe2017discrete} but limits the interactions that can be represented. Energy-based models in continuous latent spaces, by contrast, typically rely on short-run MCMC~\cite{pang2020learning}, which provides only approximate samples.
General (non-restricted) Boltzmann machines allow arbitrary pairwise interactions but are classically intractable to sample from in the regimes relevant for learning: standard iterative methods require exponentially many steps to produce independent samples, making gradient estimation prohibitively expensive as the system size increases.

Quantum annealing \cite{kadowaki1998quantum} provides a potential route beyond this classical barrier, as the hardware natively implements general Ising Hamiltonians and can, in principle, sample from non-restricted Boltzmann machines with arbitrary connectivity without imposing structural constraints on the prior~\cite{DQApaper}.
Indeed, quantum annealing hardware has been employed as a sampler for training Boltzmann machines \cite{Vinci2019Path,VuffrayPRX2022,NelsonPRAppl2022}. However, most existing approaches use slow annealing schedules---originally designed for ground-state search \cite{born1928adiabatic,farhi2000quantum}---and fit an effective inverse temperature a posteriori, treating the annealer output as a Boltzmann distribution at an unknown temperature. Under slow annealing, the output distribution is not guaranteed to follow a Boltzmann form in the first place, so fitting an effective temperature to a potentially non-Boltzmann distribution lacks a principled justification. Even setting this aside, the fitted temperature must be re-estimated at every training epoch as the model parameters evolve, incurring additional computational overhead and introducing sensitivity to the fitting procedure. Realizing the full potential of energy-based priors therefore requires a direct and principled connection between the annealing dynamics and the resulting sampling distribution.
% Realizing the full potential of energy-based priors---controlling the effective temperature for sampling concentration, applying external fields for conditional steering---requires a principled connection between the annealing dynamics and the resulting sampling distribution.

Recent theoretical analysis resolves this issue by establishing a direct connection between annealing dynamics and sampling distributions.
In the diabatic regime, the leading-order contribution of the energy dominates the output distribution, yielding samples that are well approximated by a Boltzmann form, with an explicit relationship between the annealing schedule and an effective inverse temperature \cite{DQApaper,kim2026diabatic}.
This provides a principled foundation for controlling sampling behavior through the annealing schedule, rather than relying on empirical temperature estimation.

% Recent theoretical analysis has established precisely this connection.
% When quantum annealing is operated in the diabatic limit, the leading-order contribution of the energy dominates the sampling distribution, yielding samples that are well-approximated by a Boltzmann form \cite{DQApaper,kim2026diabatic}.
% Crucially, this analysis establishes an explicit relationship between the annealing schedule and an effective inverse temperature, providing a principled basis for Boltzmann sampling.

Building on this insight, we develop variational autoencoders with Boltzmann machine priors in which the annealing schedule and external bias fields are systematically adapted to distinct stages of learning and deployment---training, unconditional generation, conditional generation, and semantic editing---within a single model. During training, quantum annealing provides samples for learning the energy-based prior; after training, the learned energy landscape is reused for unconditional generation and steered with bias fields for conditional generation and semantic editing. This multi-mode use of quantum annealing within one model is a central feature of the framework.

We demonstrate the proposed approach on CelebA \cite{liu2015deep}, a large-scale RGB dataset with rich semantic attributes, using native hardware embedding on the Zephyr topology \cite{zephyr} of a D-Wave Advantage2 processor with up to 2000 qubits, where each latent variable is mapped one-to-one to a physical qubit. The model achieves high-quality unconditional and conditional generation, showing that expressive, non-restricted Boltzmann priors can be trained and deployed effectively at scale.
Furthermore, we show that the learned energy function provides a meaningful signal for out-of-distribution detection on both MNIST and financial market data, highlighting that the energy-based prior is useful not only for generation but also for capturing informative structure relevant to downstream inference tasks.

Taken together, our results reposition quantum annealing from a black-box heuristic to a controllable computational primitive for learning, sampling, and steering structured latent energy landscapes. By combining general Boltzmann priors with quantum annealing tailored to each task, we show that energy-based latent distributions can be trained and deployed effectively within modern VAEs, enabling expressive unconditional and conditional generation as well as informative out-of-distribution detection at scales and complexities beyond prior work.

%%%%%%%%%%%%%%%%%%%%%%%%%%%%%%%%%%%%%%%%%%%%%%%%%%%%%%%%%%%
\section*{Results}
%%%%%%%%%%%%%%%%%%%%%%%%%%%%%%%%%%%%%%%%%%%%%%%%%%%%%%%%%%%

We first introduce the theoretical framework of variational autoencoders with Boltzmann-machine priors and the multi-mode quantum annealing strategy, then present experimental results on generative modeling and out-of-distribution detection.

%%% Theoretical framework %%%
\subsection*{Variational autoencoders with Boltzmann priors}\label{sec:bm_vae}

\begin{figure}[t]
\centering
\includegraphics[width=0.5\columnwidth]{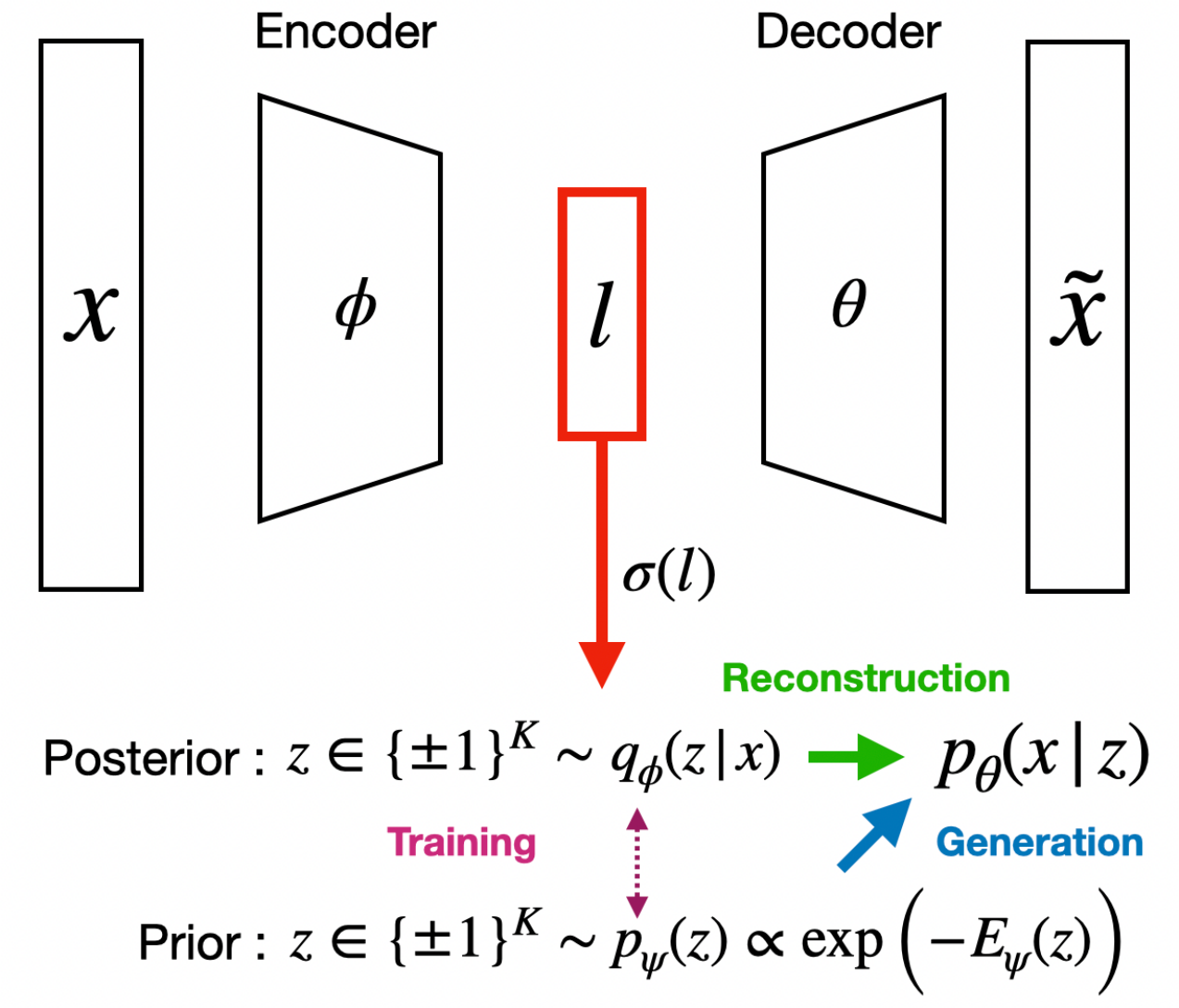}
\caption{\label{fig:BM-VAE}
Schematic illustration of a variational autoencoder with a Boltzmann prior. The encoder maps an input $x$ to a logit vector $l$. Applying the sigmoid function $\sigma$ componentwise converts these logits into the Bernoulli probabilities that parameterize the approximate posterior $q_\phi(z|x)$ over binary latent variables $z\in\{\pm1\}^K$.
% The encoder maps an input $x$ to a logit vector $l$, whose components determine the Bernoulli probabilities in the approximate posterior $q_\phi(z|x)$ over binary latent variables $z \in \{\pm 1\}^K$.
A latent sample $z$ drawn from this posterior is passed to the decoder to reconstruct $\tilde{x}$.
The Boltzmann prior $p_\psi(z)\propto \exp[-E_\psi(z)]$ is trained to match the aggregated posterior $\bar{q}(z)=\mathbb{E}_x[q_\phi(z|x)]$. After training, new samples can be generated by drawing $z$ from the learned prior and decoding it through $p_\theta(x|z)$.
% Schematic illustration of a variational autoencoder with a Boltzmann prior.
% The encoder maps input $x$ to output $\boldsymbol{\mu}$, from which binary latent variables $z \in \{\pm 1\}^K$ are sampled through the approximate posterior $q_\phi(z|x)$.
% The decoder reconstructs $\tilde{x}$ from $z$.
% The Boltzmann prior $p_\psi(z)$ is trained to match the aggregated posterior $\bar{q}(z) = \mathbb{E}_x[q_\phi(z|x)]$, enabling generation by sampling from the learned energy-based distribution.
}
\end{figure}

Our model consists of three components: an encoder $q_\phi(z|x)$, which serves as the approximate posterior and is parameterized by $\phi$; a decoder $p_\theta(x|z)$, which defines the likelihood and is parameterized by $\theta$; and a prior $p_\psi(z)$ over latent variables, parameterized by $\psi$.
Figure~\ref{fig:BM-VAE} summarizes the overall architecture.

In standard VAEs, the prior is typically fixed to an isotropic Gaussian $p(z)=\mathcal{N}(0,I)$, which factorizes across latent dimensions and encodes no learnable interactions.
We replace this factorized prior with an energy-based prior that defines the latent distribution implicitly through an unnormalized density,
\begin{equation}\label{eq:prior}
p_\psi(z) \propto \exp\!\left(-E_\psi(z)\right),
\end{equation}
where $E_\psi(z)$ is the energy function of a Boltzmann machine.
Unlike factorized priors, this formulation allows dependencies between latent variables to be represented directly through the structure of the energy function, so that the prior assigns relative plausibility to latent configurations based on their joint structure rather than serving solely as a regularizer toward a fixed reference distribution.

Training is performed by maximizing the evidence lower bound (ELBO),
\begin{equation}\label{eq:elbo}
\mathcal{L}(\theta,\phi,\psi)
=
\underbrace{\mathbb{E}_{q_\phi(z|x)}\!\left[\log p_\theta(x|z)\right]}_{\text{reconstruction}}
-
\underbrace{D_{\mathrm{KL}}\!\left(q_\phi(z|x)\,\|\,p_\psi(z)\right)}_{\text{prior matching}}.
\end{equation}
With a Boltzmann prior, the KL divergence admits a decomposition into physically interpretable components,
\begin{equation}\label{eq:kl_decomp}
D_{\mathrm{KL}}\!\left(q_\phi(z|x)\,\|\,p_\psi(z)\right)
=
\underbrace{\mathbb{E}_{q_\phi(z|x)}\!\left[E_\psi(z)\right]}_{\text{energy}}
+
\log Z_\psi
-
\underbrace{S\!\left(q_\phi(z|x)\right)}_{\text{entropy}},
\end{equation}
where $Z_\psi$ is the partition function and $S(\cdot)$ denotes entropy.
% This decomposition mirrors the structure of a variational free energy in statistical mechanics: $\langle E\rangle - S$ is the variational Helmholtz free energy $F_q$ of the posterior (at unit temperature), while $-\log Z_\psi$ is the equilibrium free energy $F_p$ of the prior.
This decomposition admits a natural interpretation in terms of statistical mechanics: $\langle E\rangle - S$ corresponds to the variational Helmholtz free energy $F_q$ of the posterior at unit temperature, while $-\log Z_\psi$ is the equilibrium free energy $F_p$ of the prior.
The KL divergence therefore equals the free-energy gap $F_q - F_p$, and minimizing it drives the posterior toward thermodynamic equilibrium with the energy-based prior.
In this view, training is not a one-sided fit to a fixed target but a \emph{joint equilibration} of two coupled systems: the encoder seeks low-energy latent configurations under the current prior while retaining sufficient entropy to support a diverse and informative representation, and the Boltzmann prior in turn reshapes its own energy landscape so that its equilibrium distribution accommodates the encoded posterior.
Learning therefore converges not when the encoder matches a prescribed distribution, but when the free energies $F_q$ and $F_p$ become mutually compatible.
A broader appeal of this energy-based formulation is that its central quantities---the energy function, its partition function, and the associated free energy---are among the most thoroughly studied objects in statistical physics.
Adopting such a prior therefore lets the framework inherit a mature toolkit built around them: Markov-chain and annealing-based samplers, free-energy estimators, and systematic approximation schemes such as mean-field, Bethe, and cluster expansions.

The expected energy and posterior entropy can be evaluated using samples from the encoder alone, but the gradient of the partition function requires expectations under the model distribution $p_\psi(z)$.
Although the ELBO in Eq.~\eqref{eq:elbo} features a per-sample KL divergence $D_{\mathrm{KL}}(q_\phi(z|x)\,\|\,p_\psi(z))$, its expectation over the data distribution admits the identity
\begin{equation}\label{eq:agg_kl}
\mathbb{E}_x\!\left[D_{\mathrm{KL}}(q_\phi(z|x)\,\|\,p_\psi(z))\right] \;=\; D_{\mathrm{KL}}(\bar{q}(z)\,\|\,p_\psi(z)) \;+\; I_{q_\phi}(x;z),
\end{equation}
where $\bar{q}(z) = \mathbb{E}_x[q_\phi(z|x)]$ is the aggregated posterior and $I_{q_\phi}(x;z)$ denotes the mutual information between $x$ and $z$ under the encoding distribution $p(x)q_\phi(z|x)$.
Because the mutual information term is independent of the prior parameters $\psi$, the prior-facing part of the training objective reduces to $D_{\mathrm{KL}}(\bar{q}\,\|\,p_\psi)$: the Boltzmann prior is effectively trained to match the aggregated posterior, as depicted in Fig.~\ref{fig:BM-VAE}.
Taking the gradient with respect to the prior parameters $\psi$ yields a positive--negative phase structure,
\begin{align}\label{eq:prior_grad}
&\nabla_\psi
\Big(
\mathbb{E}_{q_\phi(z|x)}[E_\psi(z)] + \log Z_\psi
\Big) \notag \\
&\quad=
\underbrace{\mathbb{E}_{q_\phi(z|x)}[\nabla_\psi E_\psi(z)]}_{\text{positive phase}}
-
\underbrace{\mathbb{E}_{p_\psi(z)}[\nabla_\psi E_\psi(z)]}_{\text{negative phase}},
\end{align}
which is characteristic of Boltzmann machine learning~\cite{Hinton2002}. The positive phase lowers the energy of latent configurations favored by the encoder, whereas the negative phase enforces global normalization by penalizing configurations that are overly favored by the current model.
Because the negative-phase expectation must be approximated using samples from $p_\psi(z)$, the choice of sampler becomes an integral part of the learning dynamics rather than a secondary implementation detail (see Methods for the full training objective, gradient estimators, and optimization procedure).

%%% Multi-mode sampling %%%
\subsection*{Multi-mode sampling with quantum annealing}\label{sec:tri_mode}

\begin{figure}[t]
\centering
\includegraphics[width=\columnwidth]{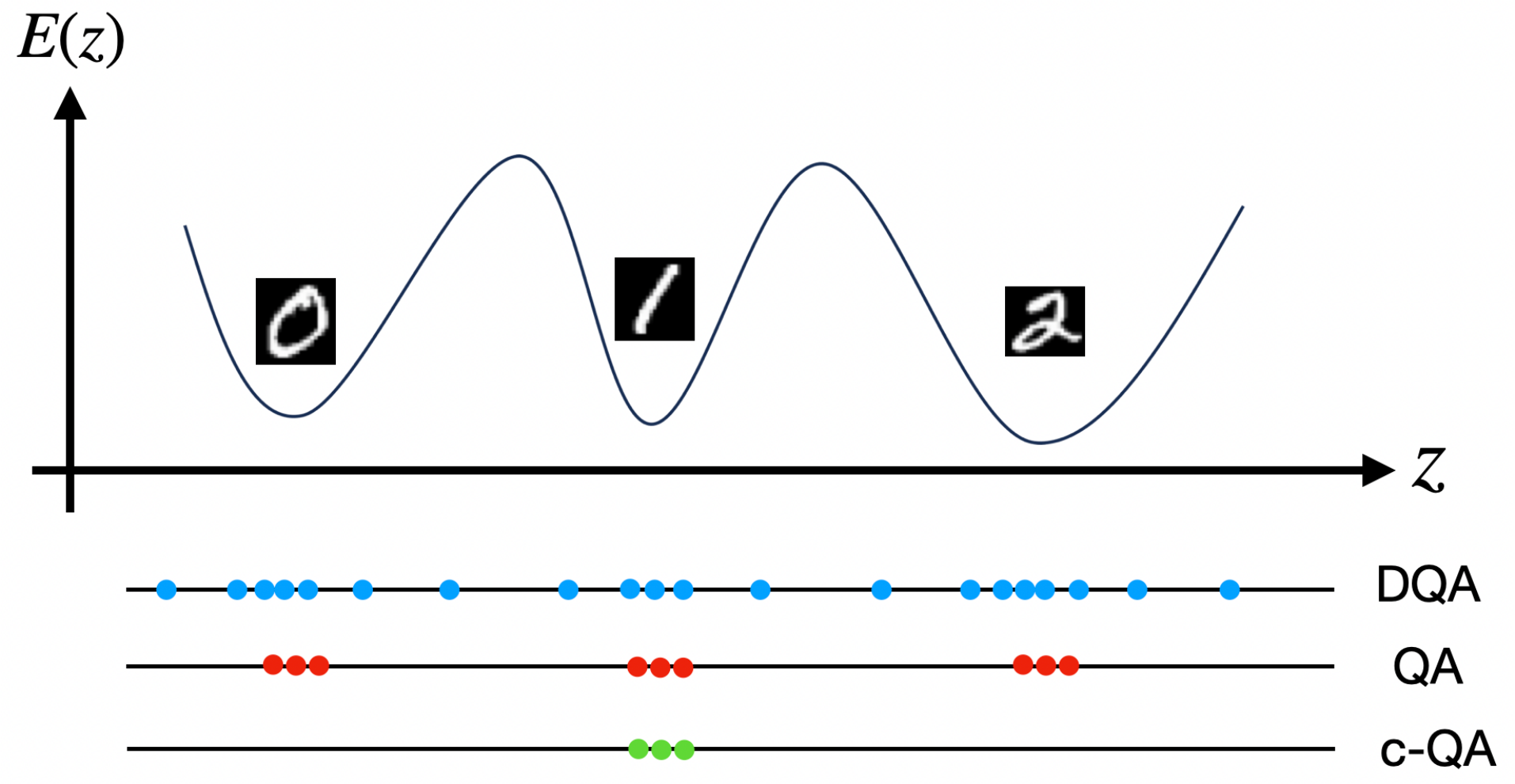}
\caption{\label{fig:sampling}Three quantum annealing modes applied to the same learned energy landscape.
Blue (DQA): diabatic quantum annealing yields samples that approximately follow a Boltzmann distribution over the landscape and are used for gradient estimation during training.
Red (QA): slower quantum annealing localizes samples near low-energy minima for unconditional generation.
Green (c-QA): conditional quantum annealing with external bias fields steers sampling toward a specific low-energy region associated with a desired attribute.
See Figs.~\ref{fig:unconditional} and~\ref{fig:conditional} for generated samples.
}
\end{figure}

The gradient update in Eq.~\eqref{eq:prior_grad} requires samples from the Boltzmann distribution defined by the current prior parameters, whereas generation from the learned energy landscape proceeds differently depending on whether conditioning is applied.
We address these distinct requirements with three quantum annealing modes operating on the same energy function (Fig.~\ref{fig:sampling}; see Methods for details).
In the diabatic regime, the annealing schedule determines an effective inverse temperature $\beta$, so that the output distribution is well-approximated by the Boltzmann form $p(z) \propto e^{-\beta\, E_\psi(z)}$ \cite{DQApaper,kim2026diabatic}. As the schedule becomes slower, the approximation to an exact Boltzmann distribution deteriorates, but the output increasingly concentrates near low-energy configurations. Accordingly, DQA (Mode~1) uses $\beta \simeq 1$ to provide samples for unbiased gradient estimation during training.
QA (Mode~2) uses slower annealing to localize samples near low-energy minima for unconditional generation.
Finally, c-QA (Mode~3) augments Mode~2 with external bias fields to steer sampling toward attribute-specific regions for conditional generation.
All three modes operate on the same Boltzmann machine without retraining.

%%% Training convergence %%%

\subsection*{Training convergence}

\begin{figure}[t]
\centering
\includegraphics[width=\textwidth]{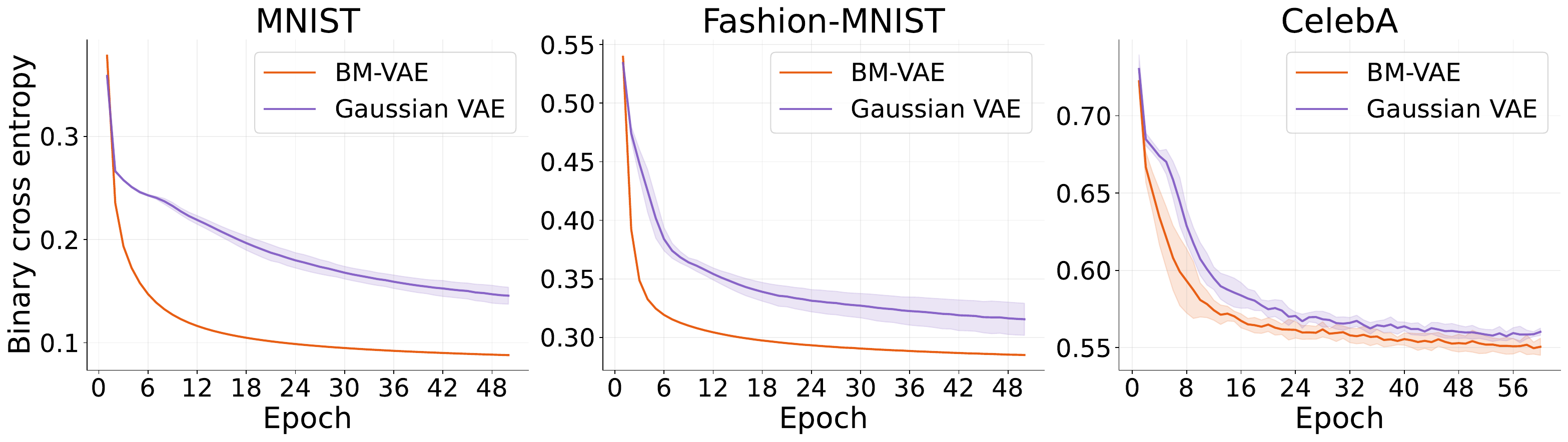}
\caption{\label{fig:learning_curves}Training curves of BM-VAE and Gaussian-prior VAE (G-VAE) on MNIST (left), Fashion-MNIST (center), and CelebA (right).
The vertical axis shows the binary cross-entropy reconstruction loss [Eq.~(\ref{eq:bce})]. Solid lines represent the mean over 10 independent runs and shaded regions indicate one standard deviation, where run-to-run variability reflects random parameter initialization and the stochastic nature of quantum annealing samples.
For each dataset, the two models share the same encoder--decoder architecture and latent dimensionality and differ only in the choice of latent prior.
}
\end{figure}

Figure~\ref{fig:learning_curves} compares the training dynamics of BM-VAE and a Gaussian-prior VAE (G-VAE) that shares the same encoder--decoder architecture on MNIST \cite{lecun2010mnist}, Fashion-MNIST \cite{xiao2017fashionmnist}, and CelebA \cite{liu2015deep} (see Methods for the model architecture).
Across all three datasets, BM-VAE converges faster and attains a lower reconstruction loss than G-VAE.
Because the Boltzmann prior is learnable, it can adapt to the encoder's output distribution rather than imposing a fixed structure, reducing the tension between reconstruction and prior matching that limits the Gaussian-prior baseline.

%%% Unconditional generation %%%

\subsection*{Unconditional generation}

In unconditional generation, new samples are produced without reference to any input data: the model must draw latent configurations entirely from the prior and decode them into realistic outputs.
A factorized prior such as $\mathcal{N}(0,I)$ samples each latent dimension independently, lacking the structured interactions needed to concentrate samples in semantically meaningful regions of the latent space.
The Boltzmann prior addresses this by encoding pairwise interactions that enforce consistency across latent dimensions, defining an energy-based distribution from which new latent configurations can be sampled via quantum annealing (Mode~2) and passed directly to the decoder.

\begin{figure}[t]
\centering
\includegraphics[width=\textwidth]{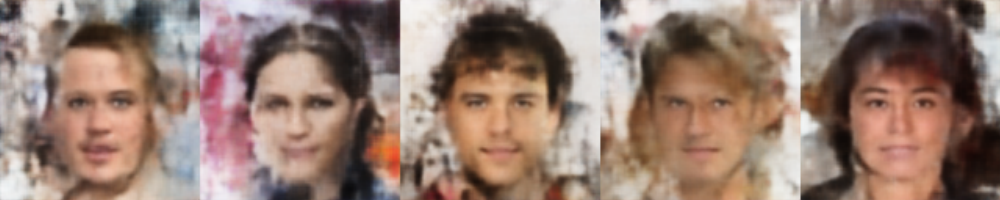}
\caption{\label{fig:unconditional}Unconditional samples from the learned Boltzmann prior on CelebA ($128\times128$, $K=2000$ latent variables).
Samples are generated on the D-Wave Advantage2 processor using QA (Mode~2), which localizes sampling near low-energy minima of the learned energy landscape.
No additional denoising or post-processing is applied.
}
\end{figure}

Figure~\ref{fig:unconditional} shows unconditional samples generated by the D-Wave Advantage2 quantum annealer, where each of the $K=2000$ latent variables is mapped one-to-one to a physical qubit on the Zephyr topology. Using QA (Mode~2), sampling concentrates near the low-energy minima of the learned energy landscape (see Methods for details of the annealing protocol).
The results reveal that diverse face configurations---varying in pose, expression, hair, and skin tone---are encoded as distinct low-energy states of the prior, confirming that the Boltzmann machine has learned a meaningful and structured latent distribution.

%%% Conditional generation %%%

\subsection*{Conditional generation via latent biasing}

% Conditional generation is performed by adding condition-dependent bias fields to the energy function of the Boltzmann prior, while keeping the encoder and decoder fixed.
While unconditional generation tests whether the learned prior captures the overall data distribution, conditional generation is more practically useful because it enables targeted synthesis of samples with desired attributes. In our framework, this is achieved by introducing condition-dependent bias fields into the energy function of the Boltzmann prior, while keeping the encoder and decoder fixed.
Physically, these bias fields act as external longitudinal fields that tilt the learned energy landscape toward configurations associated with the target attribute, so that the same annealing procedure relaxes into attribute-specific low-energy regions rather than into the unconditioned minima.

\begin{figure}[t]
\centering
\includegraphics[width=\textwidth]{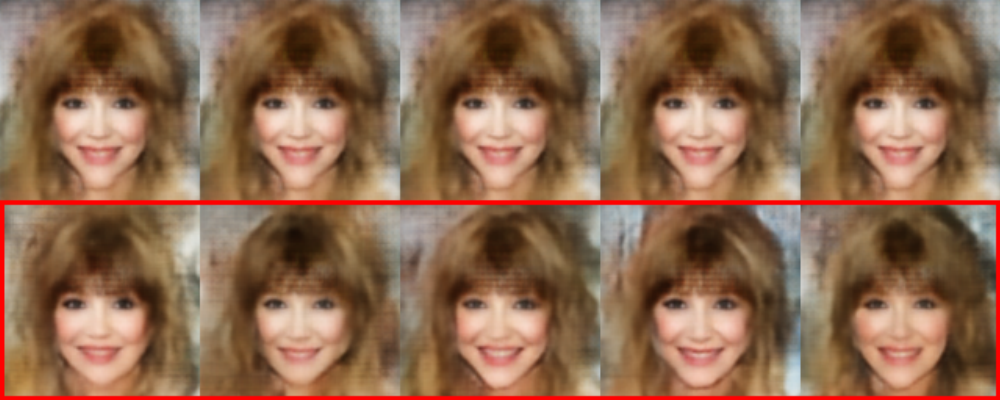}
\caption{\label{fig:conditional}Conditional generation on CelebA using the attribute-average encoder output for Bangs.
Row~1: the binarized encoder output $\mathrm{sign}(\boldsymbol{l})$ is decoded directly without quantum annealing, producing a single deterministic but visually rigid output.
Row~2: c-QA (Mode~3) with the learned couplings $J$ and bias fields $h$ derived from $\boldsymbol{l}_{\mathrm{attr}}$. The pairwise interactions of the Boltzmann prior propagate the attribute bias across latent variables, yielding samples that are both diverse and visually consistent.
}
\end{figure}

To illustrate the contribution of the learned prior, we compare two generation methods using the attribute-average encoder output for Bangs (Fig.~\ref{fig:conditional}).

The first baseline directly decodes the binarized attribute-average encoder output $z = \mathrm{sign}(\boldsymbol{l}_{\mathrm{attr}})$ without involving the prior or quantum annealing, producing a single deterministic output that appears visually rigid and unnatural.
In contrast, the second approach applies c-QA with the learned pairwise couplings $J$ of the Boltzmann prior and external bias fields $h$ derived from the attribute-average encoder output $\boldsymbol{l}_{\mathrm{attr}}$ (see Methods). The pairwise interactions of the Boltzmann prior shape the conditional sampling process by propagating the attribute bias across latent variables.
As a result, the generated samples are both diverse and semantically coherent.
This comparison demonstrates that the learned Boltzmann prior provides essential structure for high-quality conditional generation.

\begin{figure}[t]
\centering
\includegraphics[width=\textwidth]{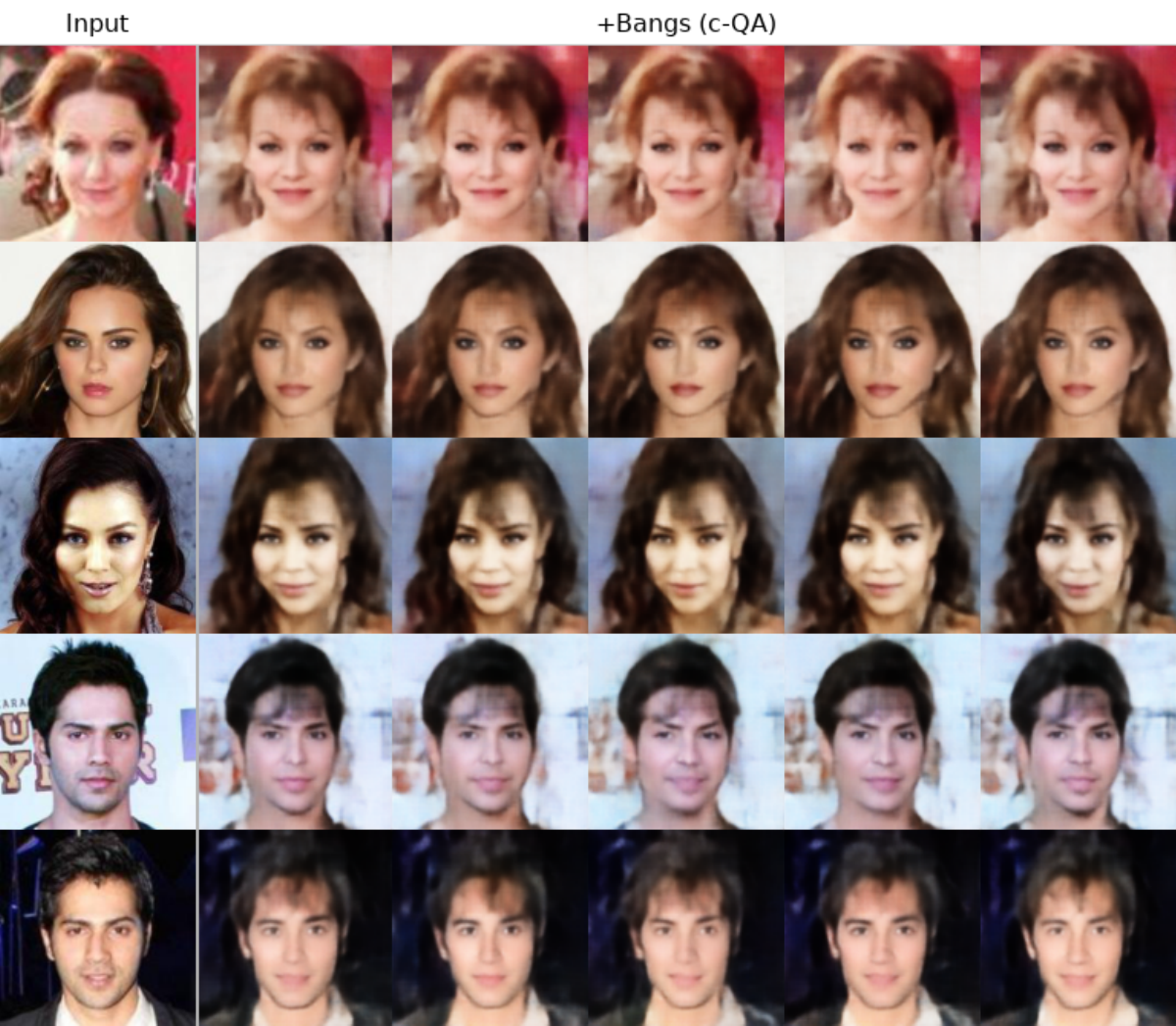}
\caption{\label{fig:attr_manipulation}Attribute manipulation via c-QA (Mode~3) on CelebA.
Left column: original test image. Remaining columns: five independent c-QA samples with Bangs added.
For each target attribute, we construct a conditioned logit vector by summing the encoder output of the test image with the attribute-average encoder output. The resulting vector defines the bias fields $h$ used in c-QA together with the learned couplings $J$. The learned prior produces semantically consistent edits while preserving the identity of the original face, with stochastic diversity across samples.
}
\end{figure}

The same mechanism also enables semantic editing of individual images.
Given a test image $x$ without a target attribute, we construct a conditioned logit vector by combining the encoder logit output of the test image with the attribute-specific direction $\bar{l}_{\mathrm{attr}} - \bar{l}$ (the attribute average minus the global average), and use the resulting bias field for c-QA (see Methods for the precise formulation).
Figure~\ref{fig:attr_manipulation} shows examples where Bangs are added to test images that originally lack this attribute. The learned prior preserves the identity of the original face while consistently introducing the desired feature, with stochastic diversity across samples.

%%% Out-of-distribution detection %%%

\subsection*{Out-of-distribution detection}\label{sec:ood}

The energy function of the Boltzmann prior provides an interpretable scalar summary of each latent configuration: inputs that conform to the training distribution occupy low-energy regions, while out-of-distribution inputs are mapped to higher-energy states.
This makes the energy a natural score for detecting out-of-distribution inputs, complementary to the reconstruction loss used in standard autoencoders.
Such out-of-distribution detection can be directly applied to anomaly detection in practical settings where anomalies manifest as distributional shifts.

\begin{figure}[t]
\centering
\includegraphics[width=\textwidth]{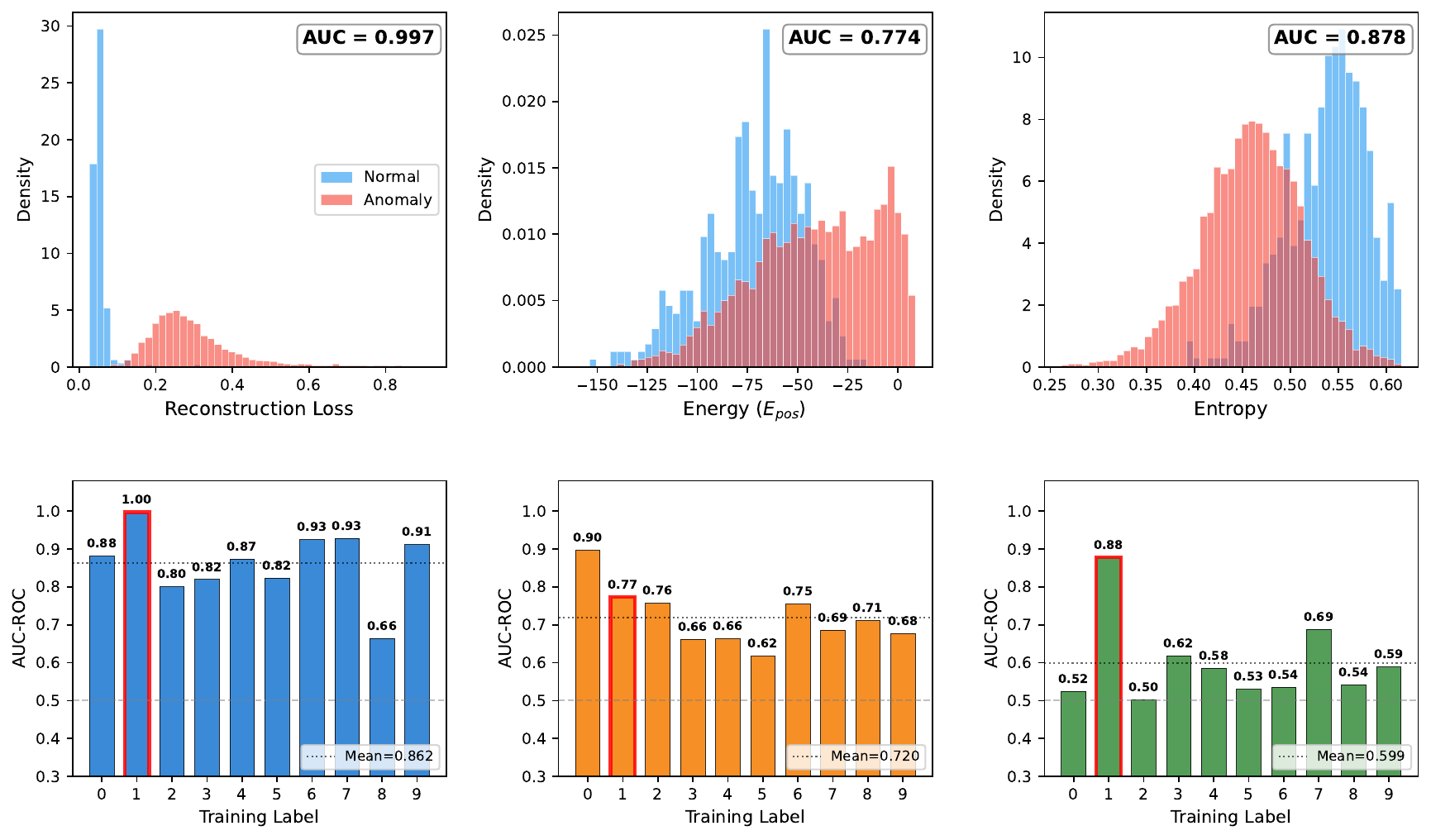}
\caption{\label{fig:mnist_anomaly}One-class anomaly detection on MNIST.
BM-VAE is trained on a single digit class (normal); all other digit classes are treated as anomalies.
Top row: score distributions of normal (blue) and anomalous (red) test samples when digit ``1'' is the normal class, shown for reconstruction loss (left), Boltzmann energy (center), and posterior entropy (right).
Bottom row: per-class AUC when each digit 0--9 serves as the normal class; the red bar highlights digit ``1'' corresponding to the top row. Dashed lines indicate the mean AUC across all ten classes.
Reconstruction loss achieves the highest mean AUC (0.862), while energy (0.720) and entropy (0.599) capture complementary aspects of distributional shift.
}
\end{figure}

\paragraph{One-class classification on images.}
We first evaluate this on MNIST using a one-class classification protocol: BM-VAE is trained on a single digit class, and all other classes are treated as out-of-distribution at test time.
Figure~\ref{fig:mnist_anomaly} shows results for each of the ten digit classes used as the in-distribution target.
Across all training labels, the mean AUC is 0.862 for reconstruction loss, 0.720 for energy, and 0.599 for entropy, suggesting that different components of the ELBO capture complementary aspects of distributional shift without task-specific supervision (see Methods for the precise definitions of each anomaly score).

\paragraph{Market regime detection.}
To test the utility of the energy score on a more challenging real-world task, we apply BM-VAE to detecting market regime shifts from equity market data.

We represent each trading day by a 784-dimensional feature vector derived from rolling pairwise correlations among 40 S\&P~500 stocks \cite{plerou2002random} (see Methods for details).
The model is trained on 1{,}364 samples from a normal market period (2012--2017) and evaluated across 2005--2024, covering seven anomalous regime episodes of varying onset speed.
Following the application of autoencoders to financial data \cite{gu2021autoencoder}, we compare three models with matched latent dimensionality $K=1000$: a standard autoencoder (AE), a Gaussian-prior VAE (G-VAE), and BM-VAE trained on D-Wave Advantage2 \cite{dwave_coherence}.

\begin{table}[t]
\centering
\caption{Market regime detection performance measured by AUC ($K=1000$).
The first two rows report overall AUC using a single score type: reconstruction loss and prior-based metric (energy for BM-VAE, KL divergence for G-VAE), respectively.
Episode-level AUC is computed using the best weighted combination of all available scores for each model.
Bold indicates best per row.}
\label{tab:regime}
\begin{tabular}{lccc}
\toprule
& \textbf{AE} & \textbf{G-VAE} & \textbf{BM-VAE} \\
\midrule
Recon only          & 0.793          & 0.775          & \textbf{0.793} \\
Prior metric        & --             & 0.744          & \textbf{0.757} \\
\midrule
\multicolumn{4}{l}{\textit{Rapid-onset episodes}} \\
\quad GFC (2008)    & \textbf{0.847} & 0.846          & 0.843 \\
\quad Flash Crash   & 0.906          & 0.899          & \textbf{0.911} \\
\quad EU Debt       & \textbf{0.951} & 0.940          & 0.950 \\
\quad China/Oil     & \textbf{0.890} & 0.843          & 0.884 \\
\quad COVID         & 0.885          & 0.871          & \textbf{0.891} \\
\quad Avg.          & \textbf{0.896} & 0.880          & \textbf{0.896} \\
\midrule
\multicolumn{4}{l}{\textit{Gradual regime shifts}} \\
\quad Fed (2018)    & 0.569          & 0.529          & \textbf{0.601} \\
\quad Inflation     & 0.574          & 0.546          & \textbf{0.609} \\
\quad Avg.          & 0.572          & 0.538          & \textbf{0.605} \\
\bottomrule
\end{tabular}
\end{table}

Table~\ref{tab:regime} compares the three models.
The prior metric of BM-VAE (0.757) exceeds that of G-VAE (0.744), indicating that the Boltzmann prior captures latent distributional structure more effectively than a factorized Gaussian prior.
On rapid-onset episodes, all three models perform comparably, with BM-VAE and AE sharing the highest average AUC (0.896).
This is expected: abrupt changes in the correlation structure produce distinctive reconstruction signatures that are sufficient for detection regardless of the prior.
The advantage of BM-VAE becomes most pronounced on gradual regime shifts.
BM-VAE achieves an average AUC of 0.605 on the Fed tightening and Inflation episodes, compared to 0.572 for AE and 0.538 for G-VAE.
This improvement is attributable to the energy score, which provides a complementary view of distributional shift that reconstruction loss does not capture.
The precise mechanism by which the energy discriminates gradual regime shifts warrants further investigation.

G-VAE performs worse than AE despite having access to a prior-based score.
At $K=1000$, the KL divergence against a factorized Gaussian prior becomes a high-dimensional quantity dominated by noise, degrading rather than improving detection performance.
In contrast, the structured pairwise interactions of the Boltzmann prior yield a more informative energy signal even at the same latent dimensionality.

\begin{figure}[t]
\centering
\includegraphics[width=\textwidth]{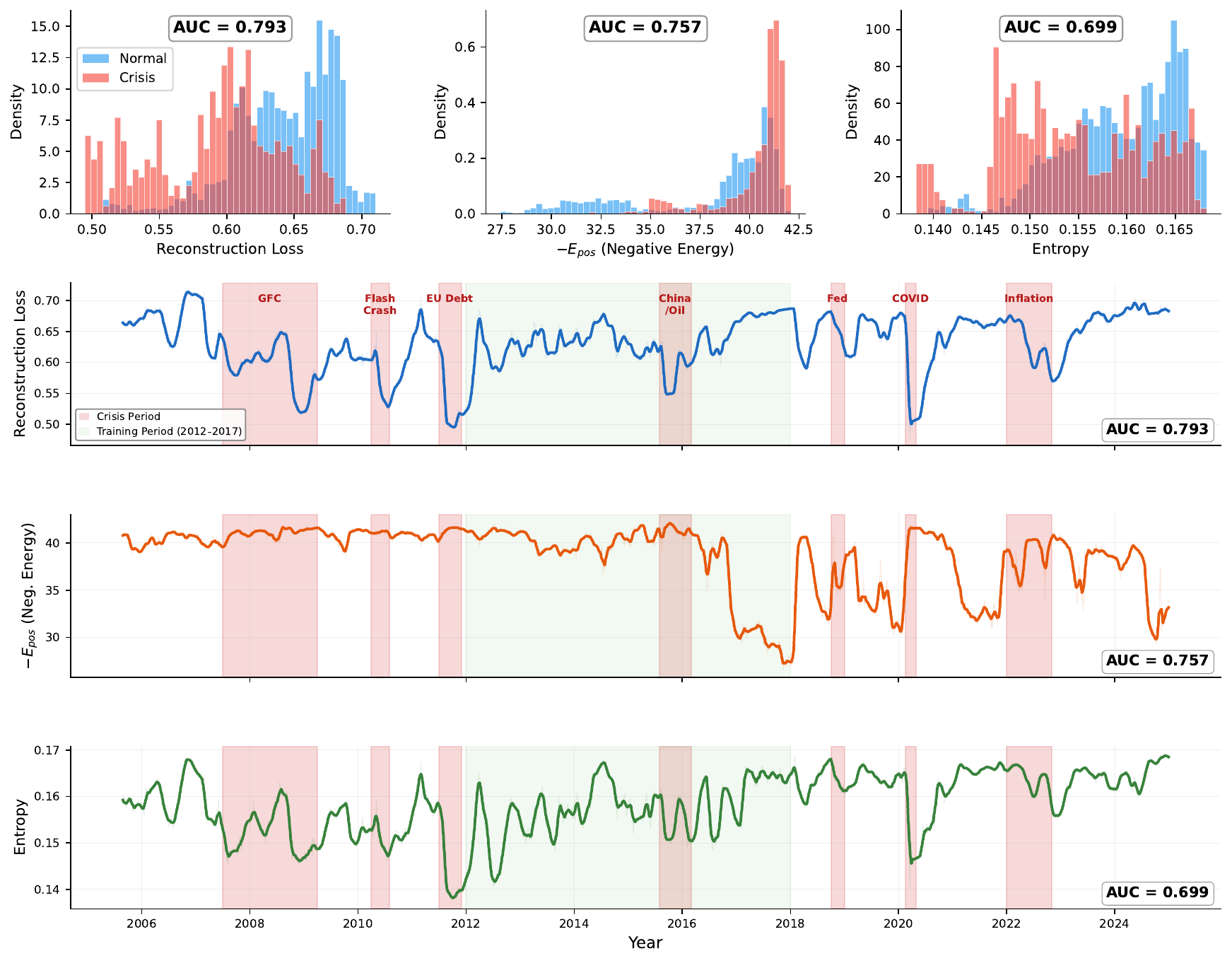}
\caption{\label{fig:financial}Score distributions and time series for market regime detection using BM-VAE ($K=1000$).
Top row: histograms of normal-period (blue) and anomalous-regime (red) samples for reconstruction loss (left, AUC 0.793), negative expected energy $-A_E$ (center, AUC 0.757), and posterior entropy (right, AUC 0.699).
Bottom rows show each score as a time series from 2005 to 2024: reconstruction loss (second row), energy (third row), and entropy (fourth row).
Red shading marks the seven anomalous regime episodes listed in Table~\ref{tab:regime} and green shading marks the training period (2012--2017).
Note that both reconstruction loss and energy are lower during anomalous regimes, reflecting the increased correlation structure of stressed markets (see text).
}
\end{figure}

Figure~\ref{fig:financial} visualizes the score distributions and time series for the three detection metrics of BM-VAE.
Reconstruction loss (AUC 0.793) serves as the primary detection signal.
Notably, during market stress the pairwise correlations among equities converge toward unity \cite{longin2001extreme}, producing correlation matrices that are closer to low-rank structure and therefore easier to reconstruct. Regime-shift samples are thus detected by lower reconstruction loss, in contrast to the image setting.
The energy score (AUC 0.757) captures a complementary signal.
The same convergence compresses the latent representation into a narrow, low-energy region of the Boltzmann prior. The regime shift is therefore signaled by \emph{lower} energy, mirroring the reconstruction inversion.
Rather than responding to individual shock events, the energy time series reflects broader market regime shifts \cite{hamilton1989new}, with sustained deviations visible during the 2018 Fed tightening and 2022--2023 inflation periods.

%%%%%%%%%%%%%%%%%%%%%%%%%%%%%%%%%%%%%%%%%%%%%%%%%%%%%%%%%%%
\section*{Discussion}
%%%%%%%%%%%%%%%%%%%%%%%%%%%%%%%%%%%%%%%%%%%%%%%%%%%%%%%%%%%
Our results show that quantum annealing can serve not merely as a heuristic sampler, but as a physically motivated, controllable, and practically useful mechanism for both training and deploying structured energy-based latent priors in variational autoencoders. By exploiting the dependence of the output distribution on the annealing schedule, the same learned Boltzmann machine can be operated in multiple modes within a single framework: diabatic quantum annealing provides samples for prior training, slower annealing enables unconditional generation by localizing sampling near low-energy minima, and conditional annealing with external bias fields supports controllable generation without retraining. This multi-mode reuse of a single learned energy landscape is a central feature of the framework, enabling practical and expressive generative modeling together with broader downstream applications such as out-of-distribution detection.

A central advance of the present work is therefore twofold. First, the latent prior is implemented as a general Boltzmann machine rather than a restricted one. Previous energy-based VAE priors have largely relied on restricted Boltzmann machines \cite{rolfe2017discrete}, whose bipartite structure is introduced primarily to enable tractable classical Gibbs sampling. Here, because quantum annealing natively implements general Ising Hamiltonians, the prior can be defined directly on the encoder output without auxiliary hidden layers or architectural restrictions imposed by classical sampling requirements. This is significant both conceptually and computationally: the learned Boltzmann machine captures pairwise interactions over an exponentially large configuration space, and training such a general fully connected prior is not scalable with standard classical sampling methods. We demonstrate that such non-restricted priors can nevertheless be trained and deployed effectively at scale, for example, on CelebA using 2000 qubits. In this sense, the present results identify a concrete regime in which quantum hardware expands the feasible design space of deep generative models.

Second, the same prior is not confined to a single role, but is trained, sampled, and manipulated through three complementary annealing modes within one model. This distinguishes the present framework from earlier QA-based VAE approaches, where annealing is used more narrowly. The generative results make the practical value of this multi-mode design clear. On CelebA, the learned Boltzmann prior supports high-quality unconditional generation, conditional generation, and semantic attribute manipulation. In particular, the comparison between direct deterministic decoding and prior-guided conditional sampling shows that the learned pairwise interactions are essential for producing samples that are both diverse and semantically coherent. Moreover, training with quantum-annealing-based sampling converges faster and to lower reconstruction loss than a Gaussian-prior VAE with the same encoder--decoder architecture. The prior therefore acts not only as a regularizer during training, but also as a reusable generative object that organizes the latent space into a structured energy landscape.

An important practical consequence is that new conditions can be imposed after training through external bias fields, without modifying the decoder or retraining the model. This supports a ``train once, condition many ways'' workflow, in which the same learned prior can be reused for unconditional generation, attribute-conditioned sampling, semantic editing, and downstream detection tasks. Such a capability may be useful in controllable content generation, scientific discovery, and inverse-design settings, where flexible navigation of a learned latent landscape is often more valuable than unconditional generation alone.

Beyond generation, the learned energy function provides an informative signal for downstream inference. Through the thermodynamic decomposition of the ELBO, the Boltzmann prior yields an energy-based score that complements reconstruction loss and posterior entropy. On MNIST, this score separates in-distribution from out-of-distribution digit classes. On financial market data, it improves the detection of market regime shifts and provides a complementary signal to reconstruction loss, particularly on gradual regime shifts. Notably, at high latent dimensionality ($K=1000$), the structured Boltzmann prior remains informative whereas the KL term of the Gaussian-prior VAE becomes substantially less useful. This indicates that the learned interactions of the Boltzmann prior capture latent distributional structure that factorized priors fail to retain.

Several directions remain open. Adaptive annealing schedules beyond the default settings used here may further improve the quality of both training and generation, and richer conditioning strategies beyond attribute-average biasing may enable more precise and compositional control over generated outputs. As quantum annealing hardware continues to improve, the framework developed here provides a natural route to deploying increasingly expressive Boltzmann priors for deep generative modeling, together with broader uses of learned latent energy landscapes for downstream inference.
\section*{Methods}\label{sec:methods}
%%%%%%%%%%%%%%%%%%%%%%%%%%%%%%%%%%%%%%%%%%%%%%%%%%%%%%%%%%%

\subsection*{Datasets}

CelebA \cite{liu2015deep} is a large-scale dataset comprising 202{,}599 aligned face images with RGB color channels.
All images are center-cropped and resized to a resolution of $128 \times 128$.
Each image is annotated with 40 binary attributes indicating the presence or absence of semantic properties such as smiling, eyeglasses, hair color, and facial hair.
Pixel intensities are normalized to $[0,1]$ and processed as three-channel RGB inputs.
Following common practice, we use the standard training, validation, and test splits provided with the dataset.

MNIST \cite{lecun2010mnist} is a dataset of 70{,}000 grayscale handwritten digit images at $28 \times 28$ resolution, normalized to $[0,1]$.
Fashion-MNIST \cite{xiao2017fashionmnist} shares the same format and resolution but contains images of clothing items across ten categories.
Both datasets are used for training convergence comparison (Fig.~\ref{fig:learning_curves}).
For one-class classification, we use MNIST.
In the one-class protocol, the model is trained on all training samples of a single digit class, and all test samples from the remaining nine classes serve as anomalies.

For market regime detection, we construct a financial market dataset from daily returns of 40 large-cap S\&P~500 stocks, sampling four stocks from each of the ten GICS sectors to ensure balanced sectoral coverage.
For each trading day, a $40 \times 40$ pairwise correlation matrix is computed from a rolling window of 60 trading days \cite{plerou2002random}.
The 780 upper-triangular entries are concatenated with four market-level summary statistics---mean and standard deviation of the pairwise correlations, average volatility, and average daily return---to form a 784-dimensional input vector.
The model is trained on 1{,}364 samples from a normal market period (2012--2017) and evaluated on all 4{,}868 samples spanning 2005--2024.
Seven anomalous regime episodes are defined and classified by onset type.
Rapid-onset episodes are those triggered by an identifiable market shock event (GFC, Flash Crash, EU Debt, China/Oil, COVID), while gradual episodes arise from prolonged macroeconomic policy shifts without a single catalytic event (Fed tightening 2018, Inflation 2022--2023).
Samples falling within these windows are excluded from the training set.

\subsection*{Model architecture}

For CelebA, we employ convolutional neural networks to capture spatial and color structure in the input images.
The encoder consists of convolutional layers followed by a fully connected layer that outputs the parameters of the approximate posterior. The number of convolutional layers is adapted to the latent dimensionality.
The decoder mirrors the encoder using transposed convolutional layers to reconstruct RGB images at resolution $128 \times 128$.
For MNIST and the financial dataset, both the encoder and decoder are single-layer fully connected networks with $K=1000$ latent variables.

The latent space is composed of $K$ binary latent variables, where $K$ denotes the latent dimensionality (see Fig.~\ref{fig:BM-VAE}), and the two latent distributions---the approximate posterior $q_\phi(z|x)$ and the prior $p_\psi(z)$---are parameterized in fundamentally different ways.
The encoder outputs an independent Bernoulli parameter for each latent variable, so that $q_\phi(z|x) = \prod_i \mathrm{Bernoulli}(z_i;\,l_i(x))$, and latent samples are obtained by sampling from this distribution during training.

The prior $p_\psi(z)$, in contrast, is modeled as a Boltzmann machine defined on the same $K$ latent variables.
Unlike the factorized posterior, the Boltzmann prior captures pairwise interactions among latent variables through its learned couplings $J_{ij}$, encoding the global structure that a factorized distribution cannot represent.
The structure of the Boltzmann prior is determined by which pairs of latent variables interact, i.e., the connectivity graph of the couplings $J_{ij}$.
On a quantum annealer, each qubit is physically connected to a fixed set of neighbors defined by the hardware topology, and only connected qubit pairs can host a nonzero coupling.
As described in the Introduction, each latent variable is mapped one-to-one to a physical qubit via native hardware embedding, and the connectivity of the Boltzmann prior directly inherits the hardware graph. We explore $K=600$ to $K=2000$ in this work.

The Boltzmann machine energy function takes the form
\begin{equation}\label{eq:energy}
E_\psi(z) = - \sum_{(i,j) \in \mathcal{E}} J_{ij} z_i z_j,
\end{equation}
where $\mathcal{E}$ denotes the set of interacting pairs determined by the hardware connectivity.
The connectivity pattern is fixed throughout training, while the coupling parameters $\{J_{ij}\}$ are learned jointly with the encoder and decoder.

\subsection*{Training objective and optimization}

Training is performed by maximizing the ELBO [Eq.~\eqref{eq:elbo}].
Each term depends on a distinct subset of parameters and is optimized using different gradient estimators, as described below.

\paragraph{Reconstruction Term.}
The reconstruction term measures the fidelity of the decoder output to the input data.
For inputs normalized to $[0,1]$, we model each output as an independent Bernoulli variable and use the binary cross-entropy (BCE) loss,
\begin{equation}\label{eq:bce}
-\log p_\theta(x|z) = -\sum_{d=1}^{D}\left[x_d \log \hat{x}_d + (1-x_d)\log(1-\hat{x}_d)\right],
\end{equation}
where $\hat{x}_d = f_\theta(z)_d$ is the $d$-th component of the decoder output and $D$ is the input dimensionality.
The reconstruction term depends on the encoder and decoder parameters $(\phi,\theta)$ through samples drawn from the approximate posterior $q_\phi(z|x)$.
Gradients with respect to the decoder parameters $\theta$ are computed analytically from the reconstruction likelihood.

\paragraph{KL Term and Encoder Gradient.}
The KL divergence term depends on the encoder parameters $\phi$ through the expected energy $\langle E_\psi(z)\rangle_{q_\phi}$ and the posterior entropy $S(q_\phi)$ [Eq.~\eqref{eq:kl_decomp}].
In practice, the encoder gradient is computed as
\begin{equation}\label{eq:encoder_grad}
\nabla_\phi \mathcal{L} = \nabla_\phi \mathbb{E}_{q_\phi(z|x)}\!\left[\log p_\theta(x|z)\right] - \lambda\, \nabla_\phi D_{\mathrm{KL}}\!\left(q_\phi(z|x)\,\|\,p_\psi(z)\right),
\end{equation}
where $\lambda \geq 0$ scales the KL contribution to the encoder gradient, independently of the prior gradient which always receives the full KL signal.
In a standard Gaussian-prior VAE, the prior is fixed, so the encoder alone must reconcile two competing objectives: preserving information for reconstruction and reshaping the posterior to match the prior.
When the prior is learnable, as in BM-VAE, the prior parameters are simultaneously updated toward the aggregated posterior through the positive--negative phase gradient [Eq.~\eqref{eq:prior_grad}], relieving the encoder of part of this burden.
The parameter $\lambda$ controls how this responsibility is shared: a smaller $\lambda$ allows the encoder to focus on reconstruction while the prior adapts to meet the posterior, whereas a larger $\lambda$ additionally drives the encoder to conform to the current prior.
This role is analogous to $\beta$ in the $\beta$-VAE framework \cite{higgins2017betavae}.
When $\lambda$ is sufficiently small, the optimization of the encoder--decoder and the prior becomes effectively decoupled: the encoder and decoder focus on reconstruction fidelity, while the Boltzmann prior captures the distributional structure of the latent space needed for generation.

\paragraph{KL Term and Prior Gradient.}
The prior parameters $\psi$ are updated using the positive--negative phase gradient derived in Eq.~\eqref{eq:prior_grad}.
The positive-phase expectation is estimated using samples from the encoder $q_\phi(z|x)$, while the negative-phase expectation requires samples from the prior $p_\psi(z)$, obtained via quantum annealing as described in the next section.

\subsection*{Quantum annealing across three modes}

The same Boltzmann machine prior is used in three distinct quantum annealing modes, each defined by a different annealing schedule and, optionally, external bias fields.

\paragraph{Mode 1: DQA for training.}
During training, the negative-phase samples required for the prior gradient [Eq.~\eqref{eq:prior_grad}] are drawn using diabatic quantum annealing (DQA) with an annealing time of 5\,ns.
Following \cite{DQApaper,kim2026diabatic}, this fast schedule yields $\beta \simeq 1$, so that the sampler reproduces the target distribution $p_\psi(z) \propto e^{-E_\psi(z)}$ without distortion, providing unbiased gradient estimates for the prior parameters.

\paragraph{Mode 2: QA for unconditional generation.}
A slower annealing schedule (0.5\,$\mu$s) concentrates samples toward low-energy configurations of the learned prior.
In the adiabatic limit, the quantum adiabatic theorem \cite{farhi2000quantum} guarantees that the system remains in its instantaneous ground state, directly yielding low-energy solutions. Even outside this limit, the diabatic framework \cite{DQApaper,kim2026diabatic} predicts the same concentration effect through an increased effective inverse temperature.
In practice, the schedule does not need to reach the true adiabatic regime. It suffices that $\beta$ is large enough to produce visually natural samples.
The sampled latent configuration $z$ is then passed through the trained decoder $f_\theta(z)$ to produce the output image.

\paragraph{Mode 3: c-QA for conditional generation.}
Conditional generation follows the same 0.5\,$\mu$s annealing procedure as Mode~2, but additionally applies external bias fields to propose desired semantic features.
This is analogous to applying an external field in an Ising model: bias fields augment the energy function, and the learned pairwise interactions $J_{ij}$ propagate these biases across latent variables, producing semantically consistent conditional samples.
We define a conditioned energy
\begin{equation}
E_{\psi,c}(z) = E_\psi(z) + E_c(z),
\end{equation}
where $E_c(z) = - \sum_i b_i(c)\, z_i$ encodes the desired condition via external bias fields $b_i(c)$.
In practice, $b_i(c)$ is constructed from encoder statistics of labeled data.
Let $l_i(x)$ denote the $i$-th component of the encoder's logit output, which parameterizes the Bernoulli posterior $q_\phi(z_i\mid x)$, and let $y(x)\in\{0,1\}$ be the ground-truth binary label of image $x$ for the target attribute $c$ (e.g., $y(x)=1$ if $x$ has \texttt{Bangs}).
Writing $\mathcal{D}^{(+)} = \{x\in\mathcal{D} : y(x)=1\}$ for the subset of images carrying the attribute, we first compute the empirical mean logit
\begin{equation}
\bar{l}_i^{(+)} \;=\; \mathbb{E}_{x \sim p(x\mid y=1)}\!\left[l_i(x)\right] \;\approx\; \frac{1}{|\mathcal{D}^{(+)}|}\sum_{x\in\mathcal{D}^{(+)}}l_i(x),
\end{equation}
and convert it into a bounded ``soft spin'' value in $[-1,1]$ via
\begin{equation}
m_i^{(+)} \;=\; \tanh\!\left(\tfrac{1}{2}\,\bar{l}_i^{(+)}\right) \;=\; 2\sigma(\bar{l}_i^{(+)}) - 1,
\end{equation}
which is the mean latent configuration characteristic of the attribute in the $\pm 1$ convention.
We then use $m_i^{(+)}$ directly as the bias direction,
\begin{equation}
b_i(c) = \gamma\, m_i^{(+)},
\end{equation}
where $\gamma$ controls the strength of the conditioning.
This biases the sampler toward latent configurations characteristic of the target attribute: dimensions where $m_i^{(+)}$ is large in magnitude receive strong bias, while those near zero are left largely unaffected.
For multi-attribute conditions, the biases are combined additively across attributes.
The goal is not to obtain equilibrium Boltzmann samples from the conditioned distribution, but to bias the sampler toward low-energy regions of the conditioned energy landscape $E_{\psi,c}(z)$, prioritizing semantic consistency with the conditioning signal.
As in Mode~2, the sampled latent configuration is passed through the decoder to produce the output.

The same mechanism extends to semantic editing of individual images.
Given a test image $x$ and a target attribute $c$, the bias field for semantic editing is constructed as
\begin{equation}
b_i^{\mathrm{edit}}(x, c) \;=\; \gamma\,\tanh\!\left(\tfrac{1}{2}\left[l_i(x) + \alpha\,\bigl(\bar{l}_i^{(+)} - \bar{l}_i\bigr)\right]\right),
\end{equation}
where $l_i(x)$ is the encoder logit of the test image, $\bar{l}_i^{(+)}$ is the attribute-average logit defined above, $\bar{l}_i = \mathbb{E}_x[l_i(x)]$ is the global average logit, and $\alpha$ controls the editing strength.
Subtracting the global average $\bar{l}_i$ isolates the attribute-specific direction in logit space, so that only the components distinctive to the target attribute are injected into the bias field while preserving the identity of the original image.

\subsection*{Anomaly detection scores}

Anomaly detection is the task of identifying inputs that deviate significantly from a learned notion of normality.
Because anomalous events are rare and diverse by nature, it is often impractical to collect labeled examples of all possible anomalies.
A common strategy is therefore one-class training \cite{ruff2018deep}: a model is trained exclusively on normal data, and at test time any input that is poorly explained by the learned model is flagged as anomalous.
Generative models such as VAEs are naturally suited to this setting, since they learn an explicit representation of the normal data distribution during training \cite{an2015variational}.

\paragraph{Thermodynamic Interpretation of the KL Term.}
As shown in Eq.~\eqref{eq:kl_decomp}, the KL divergence decomposes into the expected energy $\langle E_\psi(z)\rangle_{q_\phi}$, the posterior entropy $S(q_\phi)$, and the log-partition function $\log Z_\psi$.
Since $\log Z_\psi$ depends only on the Boltzmann machine parameters $\psi$, which are fixed after training, it takes the same value for every input $x$ and cannot serve as an anomaly signal.
The remaining two input-dependent terms each carry a distinct thermodynamic meaning:
\begin{itemize}
\item \textbf{Expected energy} $\langle E_\psi(z)\rangle_{q_\phi}$: measures how compatible the encoder's latent representation is with the learned energy landscape of the prior. A normal input is encoded into a low-energy region that the prior has learned to favor, whereas an anomalous input is likely to land in a high-energy region that the prior assigns low probability.
\item \textbf{Posterior entropy} $S(q_\phi)$: reflects how much encoding burden the encoder bears versus the prior. When the Boltzmann prior is well-trained, it captures pairwise correlations among latent variables, effectively sharing the representational load. The encoder can then afford a diffuse posterior (high entropy) for familiar inputs without sacrificing reconstruction quality. For out-of-distribution inputs, the prior offers no such support, and the encoder---never having been trained to increase entropy for these inputs---produces a concentrated posterior (low entropy). This makes entropy a complementary anomaly signal.
\end{itemize}
Together with the reconstruction loss, these quantities provide three independent views of how an input deviates from the training distribution, each probing a different stage of the generative pipeline.

\paragraph{Anomaly Score Candidates.}
Based on the quantities described above, we define three per-input anomaly scores:
\begin{align}
A_{\mathrm{rec}}(x) &= - \mathbb{E}_{q_\phi(z|x)}[\log p_\theta(x|z)], \label{eq:score_rec} \\
A_{E}(x) &= \mathbb{E}_{q_\phi(z|x)}[E_\psi(z)], \label{eq:score_energy} \\
A_{S}(x) &= S\!\left(q_\phi(z|x)\right), \label{eq:score_entropy}
\end{align}
corresponding to the reconstruction loss, the expected energy under the prior, and the posterior entropy, respectively.

\paragraph{Evaluation Protocol.}
Anomaly detection performance is evaluated using the area under the receiver operating characteristic curve (AUC).
The AUC measures the probability that a randomly chosen anomalous sample receives a higher anomaly score than a randomly chosen normal sample, with a value of 1.0 indicating perfect separation and 0.5 corresponding to random guessing.
All anomaly scores are computed independently for each test sample without additional post-processing.

%%%%%%%%%%%%%%%%%%%%%%%%%%%%%%%%%%%%%%%%%%%%%%%%%%%%%%%%%%%
\section*{Acknowledgements}
%%%%%%%%%%%%%%%%%%%%%%%%%%%%%%%%%%%%%%%%%%%%%%%%%%%%%%%%%%%

This work is supported by Institute of Information \& communications Technology Planning \& evaluation (IITP) grant funded by the Korea government (No. 2019-0-00003, Research and Development of Core Technologies for Programming, Running, Implementing and Validating of Fault-Tolerant Quantum Computing System), the National Research Foundation of Korea (RS-2025-02309510), the Ministry of Trade, Industry, and Energy (MOTIE), Korea, under the Industrial Innovation Infrastructure Development Project (RS-2024-00466693), and by Korean ARPA-H Project through the Korea Health Industry Development Institute (KHIDI), funded by the Ministry of Health \& Welfare, Korea (RS-2025-25456722).

\section*{Data availability}
CelebA \cite{liu2015deep}, MNIST \cite{lecun2010mnist}, and Fashion-MNIST \cite{xiao2017fashionmnist} are publicly available benchmark datasets.
Financial market data were obtained from public sources.

\section*{Code availability}
Code will be made available upon reasonable request.

\section*{Author contributions}
DKP conceived the project. GK and DKP developed the methodology. GK designed and performed experiments. GK and DKP analyzed the results and wrote the manuscript.

\section*{Competing Interests}
The authors declare no competing interests.

\bibliographystyle{unsrtnat}
\bibliography{references.bib}

\end{document}